\PassOptionsToPackage{table,xcdraw}{xcolor}

\documentclass[10pt,conference]{IEEEtran}

\usepackage{alltt                                    
	, multirow
	,subfig
	, booktabs
	, listings
	, graphicx
	,float
	,cite
	,verbatim
	,mathtools
	,url
}

\usepackage{xcolor}
\usepackage{xcolor, soul}
\usepackage[numbers]{natbib}
\usepackage{threeparttable}
\usepackage{framed}
\usepackage{expl3}
\usepackage{hyperref}
\usepackage{filecontents}

\usepackage{pgfplots}
\pgfplotsset{compat=1.16}
\usepackage{pgfplotstable}

\usepgfplotslibrary{statistics}

\usepackage{scalefnt}
\usepackage{tikz}

\usepackage{tabularx}
\usepackage{lipsum}
\usepackage{array}

\usepackage{tcolorbox}
\usepackage{multirow}

\usetikzlibrary{shapes.gates.logic.US,trees,positioning,arrows}
\usetikzlibrary{shapes.multipart}
\usetikzlibrary{shapes,arrows}

\tikzstyle{vertex}=[ellipse,fill=black!25,minimum size=20pt, inner sep=0pt]
\tikzstyle{edge} = [draw,thin,-]
\tikzstyle{glabel} = [text width=1cm,text centered,font=\bf]
\pgfdeclarelayer{bg}    
\pgfsetlayers{bg,main} 

\ExplSyntaxOn
\newcommand\latinabbrev[1]{
  \peek_meaning:NTF . {
    #1\@}%
  { \peek_catcode:NTF a {
      #1., \@ }%
    {#1., \@}}}
\ExplSyntaxOff

\usepackage{tikz}
\usepackage{verbatim}
\usetikzlibrary{arrows,shapes,backgrounds}

\tikzstyle{vertex}=[ellipse,fill=black!25,minimum size=20pt, inner sep=0pt]
\tikzstyle{edge} = [draw,thin,-]
\tikzstyle{glabel} = [text width=1cm,text centered,font=\bf]
\pgfdeclarelayer{bg}    
\pgfsetlayers{bg,main}  

\usepackage{soul}
\usepackage{pifont}
\usepackage{booktabs,makecell}
\usepackage[export]{adjustbox}
\usepackage{balance}
\usepackage{paralist}

\usepackage{pgf-pie}

\newif\ifpienumberinlegend
\pgfkeys{/number in legend/.code=
    \expandafter\let\expandafter\ifpienumberinlegend
    \csname if#1\endcsname
    \ifpienumberinlegend

    \def\beforenumber##1\afternumber{}%
    \fi,
    /number in legend/.default=true
}

\makeatletter
\pgfplotsset{
    boxplot/hide outliers/.code={
        \def\pgfplotsplothandlerboxplot@outlier{}%
    }
}
\makeatother

\newcommand{\CASE}[1]{\STATE \textbf{case} #1\textbf{:} \begin{ALC@g}}
\newcommand{\ENDCASE}{\end{ALC@g}}

\newcommand{\DEFAULT}{\STATE \textbf{default:} \begin{ALC@g}}
\newcommand{\ENDDEFAULT}{\end{ALC@g}}
\newcommand{\DEFAULTLINE}[1]{\STATE \textbf{default:} }
\let\footnotesize\scriptsize

\newsavebox{\supbox}
\newcommand{\bsup}{\begin{lrbox}{\supbox}$\tt\scriptstyle}
\newcommand{\esup}{$\end{lrbox}{}^{\usebox{\supbox}}}

\definecolor{lightpurple}{rgb}{0.8,0.8,1}
\definecolor{codebg}{RGB}{255,255,255}
\definecolor{commentcolor}{RGB}{11,140,11}
\usepackage[normalem]{ulem} 
\usepackage{xcolor}

\usepackage{ifthen}
\usepackage{amssymb}
\newboolean{showcomments}

\setboolean{showcomments}{false}

\ifthenelse{\boolean{showcomments}}
{\newcommand{\nbc}[3]{
 {\colorbox{#3}{\bfseries\sffamily\scriptsize\textcolor{white}{#1}}}
 {\textcolor{#3}{\sf\small$\blacktriangleright$\textit{#2}$\blacktriangleleft$}}
 }
}
{\newcommand{\nbc}[3]{}
 }

 \newcommand\RotText[1]{\rotatebox{90}{\parbox{2cm}{\centering#1}}}

\usepackage{amsmath}
\usepackage{enumitem}

\begin{document}


\title{Can We Identify Stack Overflow Questions Requiring Code Snippets? Investigating the Cause \& Effect of Missing Code Snippets}

\author{
\IEEEauthorblockN{Saikat Mondal }
\IEEEauthorblockA{University of Saskatchewan\\
\ saikat.mondal@usask.ca}
\and
\IEEEauthorblockN{Mohammad Masudur Rahman}
\IEEEauthorblockA{Dalhousie University\\
\ masud.rahman@dal.ca}
\and
\IEEEauthorblockN{Chanchal K. Roy}
\IEEEauthorblockA{University of Saskatchewan\\
\ chanchal.roy@usask.ca}
}








\maketitle

\begin{abstract}
On the Stack Overflow (SO) Q\&A site, users often request solutions to their code-related problems (e.g., errors, unexpected behavior). Unfortunately, they often miss required code snippets during their question submission. Such a practice could prevent their questions from getting prompt and appropriate answers. 
In this study, we conduct an empirical study investigating the cause \& effect of missing code snippets in SO questions whenever required. 
In this paper, our contributions are threefold.
First, we analyze how the presence or absence of required code snippets in SO questions affects the correlation between question types (missed code, included code after requests \& had code snippets during submission) and corresponding answer meta-data, such as the presence of an accepted answer.
According to our analysis, the chance of getting accepted answers is three times higher for questions that include required code snippets during their question submission than those that missed the code.
We also investigate the confounding factors (e.g., user reputation) that can affect questions receiving answers besides the presence or absence of required code snippets. We found that such factors do not hurt the correlation between the presence or absence of required code snippets and answer meta-data.
Second, we surveyed 64 practitioners to understand why users miss necessary code snippets.
About 60\% of them agree that users are unaware of whether their questions require any code snippets.
Third, we thus extract four text-based features (e.g., keywords, POS-based patterns) and build six Machine Learning (ML) models to identify the questions that need code snippets. 
Our models can predict the target questions with 86.5\% precision, 90.8\% recall, 85.3\% F1-score, and 85.2\% overall accuracy, which are highly promising. 
Our work has the potential to (a) save significant time in programming question-answering and (b) improve the quality of the valuable knowledge base by decreasing unanswered and unresolved questions. 

\end{abstract}


\section{Introduction}
\label{sec:introduction}

Stack Overflow (SO) is the largest online programming-related Q\&A site. More than 5.5K questions are posted on SO daily, and such a number is increasing rapidly \cite{datadumpapi, emse2021bmondal}. A large number of questions discuss the programming problems (e.g., coding errors, unexpected behavior) that warrant code snippets for a resolution \cite{treude2011programmers, squire2014bit}. 
Traditionally, users at SO first analyze the code snippets to identify or reproduce the reported problems \cite{emse2021bmondal}. Upon success, they can submit appropriate solutions. Unfortunately, question submitters often underestimate the necessity of code snippets or simply miss the required code snippets, which might prevent these questions from getting appropriate answers in a timely fashion. Such a scenario might also explain the 31\%  unanswered and more than 50\% unresolved questions at SO \cite{datadumpapi,rahman2015insight,asaduzzaman2013answering}. 
Interestingly, SO allows discussion where users can request code snippets using comments. However, question submitters might not always be able to respond on time due to various factors. For example, they might be connected to SO from different time zones, which could lead to unexpected delays in question-answering. Given all these 
challenges in question answering, a comprehensive understanding of why question submitters miss required code snippets, and how missing code snippets affect the questions at SO is 
warranted.


Several existing studies investigate whether the presence of code snippets in SO questions improves their quality or not \cite{asaduzzaman2013answering, squire2014bit, calefato2018ask, wang2018understanding, neshati2017early, baltadzhieva2015predicting, chua2015answers, yazdaninia2021characterization, qclassification, treude2011programmers}. 
However, the existing findings are quite mixed. According to the literature, code snippets have both positive \cite{calefato2018ask, qclassification, asaduzzaman2013answering, treude2011programmers} and negative \cite{chua2015answers, baltadzhieva2015predicting}  effects on questions.
For example, Calefato et al.~\cite{calefato2018ask} suggest that the presence of code snippets in questions increases their chance of getting appropriate answers.
On the contrary, Baltadzhieva et al.~\cite{baltadzhieva2015predicting} estimate the influence of code snippets on questions and argue that the presence of code snippets might hurt the questions' chance of getting answers.
In particular, redundant and complex code snippets in a question could increase its analysis overhead \cite{mondal2022reproducibility, baltadzhieva2015predicting}.
On the other hand, questions requiring but missing code snippets might not be answered on time.
Existing studies could not reach a consensus and thus warrant further investigation on the effect of code snippets on SO questions.


\begin{figure}[!htb]
	\centering
	\includegraphics[width = 3.3in]{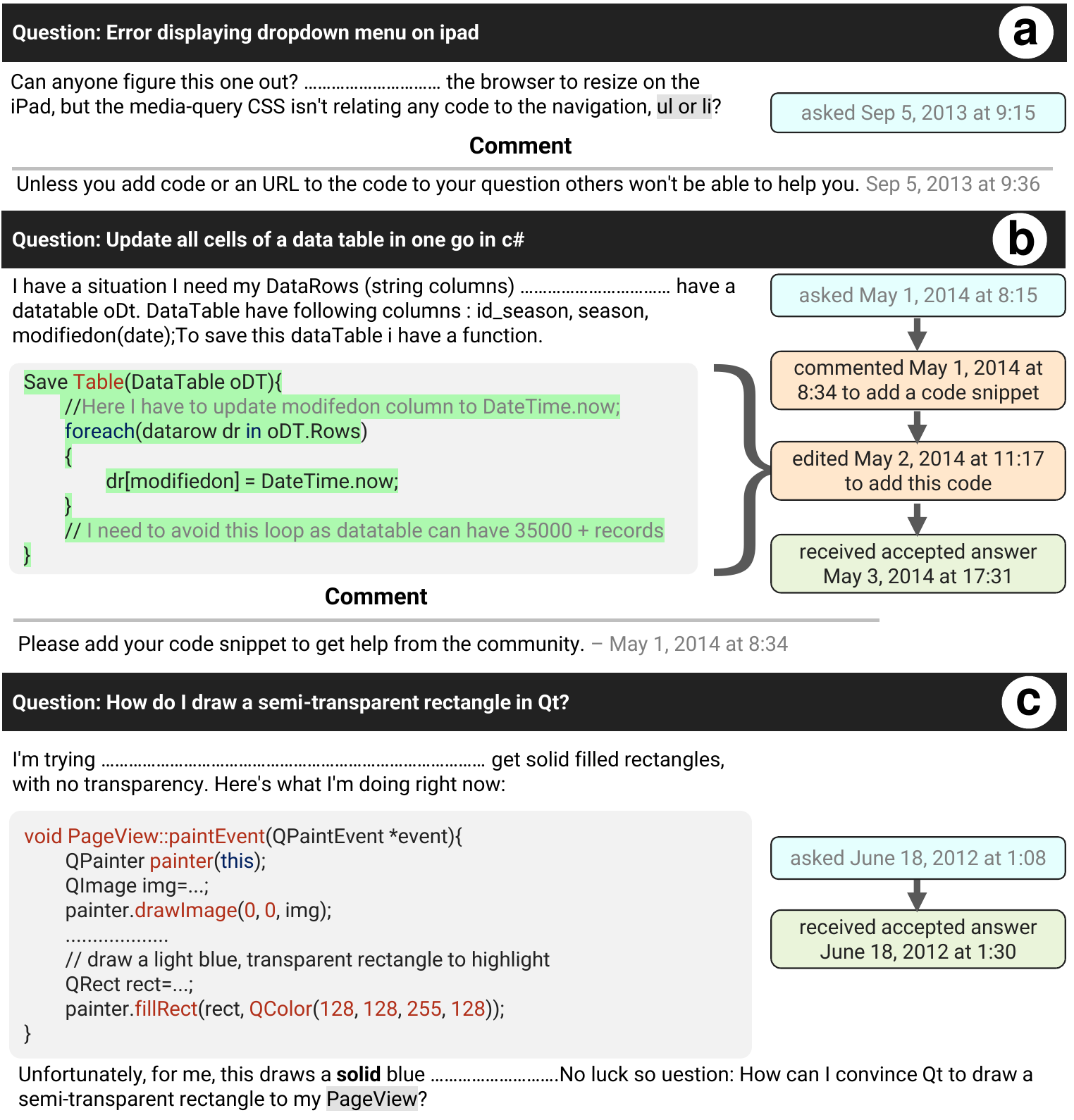}
	\caption{Motivating examples}
	\label{fig:motivationExamples}
	\vspace{-5mm}
\end{figure}

In this paper, we conduct an empirical study to (a) analyze the impact of missing code snippets (when required) in SO questions, (b) understand why users miss the required code snippets during question submission, and (c) automatically detect questions that need code snippets using machine learning.
First, we randomly select 1,207 SO questions that need code snippets (400 missed code snippets + 400 included code after requests + 407 had code snippets during submission) by applying a few heuristics. We also manually validate all these questions to avoid any false positives. Then, we show the empirical evidence suggesting that the absence of required code snippets could prevent the questions from getting their answers.
We also investigate the confounding factors such as \emph{user reputation} and \emph{question submission time} besides the presence or absence of required code snippets and see their effect on questions receiving answers.
Second, we survey 64 practitioners (e.g., SO users) to gather further insights into why the question submitters might miss the required code snippets in their questions.
Third, we randomly select 1,207 questions that do not need code snippets by applying a few heuristics.
We then pick 70\% (845 out of 1,207) of questions from each category (questions that need code \& do not need code). Next, we conduct a comparative analysis between two categories of questions to extract features that can classify them. Our comparative analysis extracts four text-based features.
We then develop six ML models (e.g., Random Forest, SVM) using these features to identify questions that need code snippets automatically.
In particular, we answer three research questions through three major contributions in this paper as follows:

     \noindent\textbf{RQ\textsubscript{1}: How do the answers get affected when the questions miss the required code snippets at Stack Overflow? What other factors affect questions receiving answers besides missing required code?}
     We determine the correlation between question types (missed code + included code after requests + had code during submission) and the presence of an accepted answer (\textbf{RQ\textsubscript{1}a}), time delay getting an accepted answer (\textbf{RQ\textsubscript{1}b}), and the presence of answers (\textbf{RQ\textsubscript{1}c}). According to our investigation, only 23.8\% of questions get acceptable answers that miss the required code snippets. On the contrary, such a percentage is 61.4\% for the questions that include code snippets during their submission. About 44\% of the questions that add code snippets upon users' request get acceptable answers. However, the delay in getting acceptable answers is significantly higher for the questions that either miss the code during submission or add the code based on request. 
     Furthermore, 28\% of questions that miss code snippets remain unanswered, whereas such percentage is only 8.8\% for those that include code snippets during submission. All these findings suggest that including required code snippets in questions during submission encourages answers, including acceptable answers with a minimum delay.

    We also investigate two confounding factors (user reputation \& question submission time) and see whether such factors hurt the correlation between the presence or absence of required code snippets and answer meta-data (\textbf{RQ\textsubscript{1}d}). According to our investigation, these factors might influence questions receiving accepted answers. However, regardless of these confounding factors, questions with required code snippets have a significantly higher chance of receiving accepted answers with minimum time delay and encourage more answers.

    \noindent\textbf{RQ\textsubscript{2}: Why do question submitters miss the required code snippets during question submission?}
    We surveyed 64 users of SO to understand why question submitters miss the required code snippets while posting questions. About 60\% of them agree that users are unaware of whether their questions need code snippets. Such evidence motivates an automatic prediction system to identify questions needing code snippets during submission.

    \noindent\textbf{RQ\textsubscript{3}: Can we predict questions that need code snippets during submission?}
    We extract four text-based features by analyzing our two types of questions that need \& do not need code snippets. Our careful analysis provides 127 unique keywords (e.g., code, try), 36 POS-based patterns from the title \& 86 POS-based patterns from body texts. Using these features, we developed six popular supervised ML techniques to identify the questions that need code snippets. These techniques are widely used in relevant studies  \cite{TUT-Saha-2013,UAC-Ponzanelli-2014,AII-Rahman-2015, beyer2018automatically}. Our models can identify the target questions with up to 86.5\% precision, 90.8\% recall, 85.3\% F1-score \& 85.2\% overall accuracy, which are promising.


\smallskip
\noindent\textbf{Replication Package} that contains our study datasets, features, ML model, and survey responses can be found in our online appendix \cite{replicationPackage}.


\begin{figure*}[t]
	\centering
	\includegraphics[width = 7in]{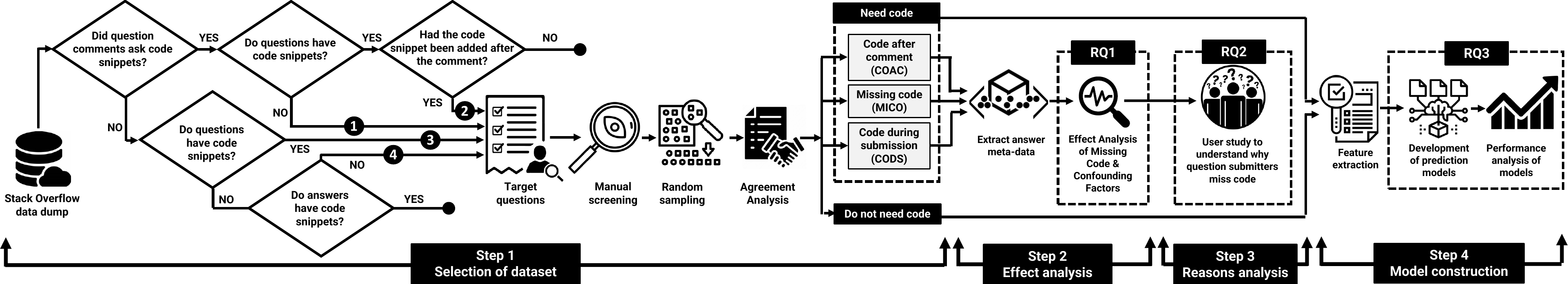}
	\caption{Study methodology}
	\label{fig:studyMethodology}
	\vspace{-4mm}
\end{figure*}


\section{Motivating Examples} 
\label{motiExample}

We present three examples
to motivate our idea regarding missing code in SO questions. The first question (Fig. \ref{fig:motivationExamples}a, \cite{motiExample1}) discusses an issue related to the display of a drop-down menu on an iPad. However, the question submitter did not include the problematic code snippet in the question. As a result, users who attempted to answer the question could not figure out the problem. One user thus commented, \emph{``unless you add code or a URL to the code to your question, others will not be able to help you''}. Unfortunately, the question submitter still did not add any code snippets, and the question did not receive any answers over the last eight years.

The second question (Fig. \ref{fig:motivationExamples}b, \cite{motiExample2}) is related to the update of all data table cells in one go. Unfortunately, it also missed the required code snippet. However, the question submitter edited the question and added an example code snippet in response to a comment. This question received an acceptable solution, but there was a long delay (e.g., more than 60 hours) in receiving the solution. 
On the other hand, the third question (Fig. \ref{fig:motivationExamples}c, \cite{motiExample3}) included the required code snippet during its submission. Interestingly, it received an acceptable solution within 22 minutes, close to the median time delay of getting an acceptable answer at SO (i.e., 21 minutes \cite{datadumpapi}).

These examples are the representatives of a large collection of similar questions that motivate us to investigate further. Thus, we plan to conduct an empirical study to (a) analyze the impact of missing but required code snippets, (b) understand why users miss them, and (c) automatically detect questions that need code snippets during their submission.

\section{Study Methodology}
\label{sec:methodology}

Fig. \ref{fig:studyMethodology} shows the schematic diagram of our study methodology. 
The following sections discuss different steps of our methodology.



\begin{table}[!htb]
	\centering
	\captionsetup{justification=centering, labelsep=newline}
	\caption{Dataset construction summary
	(\texttt{MICO} = Missing Code; \texttt{COAC} = Code After Comment;     
        \texttt{CODS} = Code During  Submission; \texttt{DONC} = Do Not Need Code)}
	\label{table:dataset-construction}
	\resizebox{3.4in}{!}{%
	
	\begin{tabular}{l|c|c|c|c}
        \hline
        \multicolumn{1}{c|}{\textbf{\begin{tabular}[c]{@{}c@{}}Questions \\ Category\end{tabular}}} & 
        \multicolumn{1}{c|}{\textbf{\begin{tabular}[c]{@{}c@{}}\# of \\ Questions\end{tabular}}} & 
        \multicolumn{1}{c|}{\textbf{\begin{tabular}[c]{@{}c@{}}Randomly \\ Sampled \\ for manual \\ analysis\end{tabular}}} & 
        \multicolumn{1}{c|}{\textbf{\begin{tabular}[c]{@{}c@{}}After \\ Removing \\ False-Positive \\ Samples\end{tabular}}} & 
        \multicolumn{1}{c}{\textbf{\begin{tabular}[c]{@{}c@{}}Randomly \\ Sampled\\  for Final \\ Analysis\end{tabular}}} \\ \hline

      \textbf{MICO}    & 24,736      & 800  & 529  & 400   \\ \hline 
      \textbf{COAC}    & 55,095      & 800  & 483  & 400  \\ \hline 
      \textbf{CODS}    & 10,244,403  & 500  & 407  & 407  \\ \hline 
      \textbf{DONC}    & 1,260,912   & 1,207 & 1,207 & 1,207   \\ \hline 
   
\end{tabular}	
}
\end{table}
\vspace{-3mm}

\subsection{Dataset Construction}
\label{subsec-mothodology:createDataset}

Fig. \ref{fig:studyMethodology} (Step 1) shows the steps of our dataset construction. We collected an August 2023 data dump of SO from the Stack Exchange site \cite{datadumpapi}.
Questions were posted in August 2023 or earlier, which suggests that SO users have assessed the questions for a significant period.

\noindent\textbf{Heuristic-based question selection.}
We aim to determine the impact of missing code snippets in the questions. We thus collect questions of three categories that need code snippets. They are questions that (1) missed code snippets (\textbf{MICO}), (2) added an entire or significant portion of code after comment (\textbf{COAC}), and (3) added code snippets during submission (\textbf{CODS}). To identify the first two categories, we look for appropriate key phrases from the question comments. We first randomly select 500 question comments whose text matched with \emph{code} keyword. Our assumption was that comments containing the term \emph{code} could request code snippets. When we manually analyze those comments, we see that the keyword code can be used to request code snippets or in different contexts. However, we found \emph{seventeen} key phrases that discuss code requirements in questions.
They are -- \emph{add code, add your code, add the code, attach code, attach your code, attach the code, include code, include your code, include the code, give your code, provide code, provide the code, provide your code, show your code, what you tried, post your code,} and \emph{post the code please.}

We then separate the questions based on whether any of their comments contain any of our key phrases or not. If one or more key phrases are matched, and the questions do not have any code segments, then we consider them as questions that missed code snippets (MICO) (Fig. \ref{fig:studyMethodology}, Step 1, label 1). We consider that these questions added code snippets after comments (COAC) (Fig. \ref{fig:studyMethodology}, Step 1, label 2) if they have code snippets. On the other hand, we assume that the questions added code segments during their submission (CODS) (Fig. \ref{fig:studyMethodology}, Step 1, label 3) if -- (a) any of the comments to questions do not match with the key phrases, and (b) questions contain code snippets. We identify the presence of code snippets in the question texts using specialized HTML tags such as \emph{$<$code$>$} under \emph{$<$pre$>$}. Then, we consider the remaining questions as potential samples for the fourth category -- questions that do not need any code snippets (DONC). We select them in Fig. \ref{fig:studyMethodology}, Step 1, label 4.

\noindent\textbf{Manual screening of false-positive samples.}
Despite the careful steps above, our data might contain false-positive samples. We thus randomly sampled 50 questions from each of the four categories and manually checked for 
false-positive samples. 
Our preliminary analysis finds that all categories except DONC contain 20\%--45\% false-positive samples, which makes them noisy.
We thus randomly sampled 800 questions from each of the two categories -- MICO \& COAC, and 500 questions from CODS for further analysis. We aim to capture 400 true-positive samples from each category. 

We involve two independent annotators to check the sampled data and discard the false positives. They are (a) the first author of this paper with 10+ years of development experience and (b) a top user of Stack Overflow with 8+ years of development experience.
We found that code snippets were added to the questions as an image file, external link, or plain text \cite{addedCodeAsImage1, addedCodeAsImage2, addedCodeAsExternalLink1, addedCodeAsExternalLink2, improperFormatCodeAfterComment}, and thus our automated analysis based on \emph{$<$code$>$} under \emph{$<$pre$>$} tags failed to identify them. 
We also carefully read all the comments to determine whether the users were requesting to add code snippets or not. We also checked the revision history of questions to ensure that code snippets were added -- (a) in response to comments and (b) during question submission. 

After manual investigation, we find 529 (out of 800) true-positive samples for MICO, 483 (out of 800) for COAC, and 407 (out of 500) for CODS. Finally, we randomly select 400 questions from each of MICO \& COAC categories and kept 407 questions from CODS category for further investigation. From the questions that do not need any code, we randomly select 1,207 questions. A total of 140 person-hours was spent on the manual investigation by the two annotators.

\begin{table*}[!htp]
\centering
    \caption{Experience, profession, and SO profile age of the participants}
	\label{table:participants-summary}
	\resizebox{5.5in}{!}{%
        \begin{tabular}{c|c|c|c|c|c|c|c|c|c|c|c|c} \hline
        \multicolumn{6}{c|}{\textbf{Development Experience (Years)}}                                                 & \multicolumn{3}{c|}{\textbf{Profession}}                                   & \multicolumn{4}{c}{\textbf{SO Profile Age (Years)}}                  \\ \hline
        \textbf{\textless{}=2} & \textbf{3-5} & \textbf{6-8} & \textbf{9-11} & \textbf{12-14} & \textbf{15 or More} & \textbf{SW Developer} & \textbf{Academician} & \textbf{Research Engineer} & \textbf{\textless{}=2} & \textbf{3-5} & \textbf{6-8} & \textbf{9-11} \\ \hline
        22   & 20   & 12    & 6    & 3   & 1    & 39   & 21  & 4    & 20  & 26  & 12   & 6  \\ \hline       
        \end{tabular}
    }
\vspace{-5mm}
\end{table*}

\noindent\textbf{Agreement analysis.}
We measure the agreement level between the two annotators above using Cohen's Kappa \cite{cohen1968weighted, cohen1960coefficient}. The value of {\large $\kappa$} was $0.99$, which means the strength of agreement is almost perfect. In particular, there were only eight disagreements (6 MICO + 2 CODS). We then send these eight samples to a third annotator, who is a senior software developer and also a top user of SO. According to the third annotator, one question needed code, but the code snippets were highly recommended for the remaining seven questions. We thus keep all eight samples in our dataset for further investigation. Interestingly, no disagreement was found in the questions that do not need any code snippets.
Note that all the sample sizes are statically significant with a 95\% confidence level and a 5\% error margin. Table \ref{table:dataset-construction} summarizes the dataset construction steps.  


\subsection{Analyzing the Effects of Code Snippets \& Confounding Factors (RQ\textsubscript{1})}
\label{subsec-mothodology:impact-analysis}

\noindent \textbf{Dataset selection for effect analysis.}
We select three categories of questions requiring code snippets: MICO, COAC \& CODS (Section \ref{subsec-mothodology:createDataset}, Table \ref{table:dataset-construction}) and determine their effect on answer meta-data.

\noindent \textbf{Recording of answer meta-data.}
To conduct our effect analysis, we record several meta-data from SO, such as the question \& accepted answer submission time and the number of answers against a question. For COAC, we record two additional items - (a) the submission time of comments requesting code snippets and (b) the addition time of code snippets to the questions. However, this information was not readily available in the SO data dump. We thus browse the comments \& revision histories of questions at the SO site and manually record them against 400 questions from the COAC category.

\noindent \textbf{Correlation analysis.}
We analyze the correlation between question types and corresponding answer meta-data, such as the presence of an accepted answer, the time delay between the submission of a question \& an accepted answer, and the presence of answers. In particular, we measure the percentage of accepted answers for each question type and determine how the presence/absence of code snippets affects the answers. It should be noted that code snippets were added to these questions at different times, which could affect their chance of getting the accepted answers. We also determine whether the percentage difference is statically significant among the questions using appropriate statistical tests (e.g., Mann-Whitney-Wilcoxon \cite{mcknight2010mann}, Cliff's Delta \cite{macbeth2011cliff}).

\noindent \textbf{Effect analysis of confounding factors.}
We investigate the correlation between question types and answer meta-data. However, a few other factors might affect questions receiving answers and hurt the correlation. In this section, we investigate the two most potential factors from literature, such as \emph{user reputation} and \emph{question submission time}, and see (a) their impact on questions receiving answers and (b) whether they hurt the correlation.


    $\bullet$ \emph{Reputation:} Stack Overflow designs a reputation system of users to incentivize their contributions \cite{calefato2015mining, mondal2021early, mondal2022reproducibility}. Several studies argue that reputations might impact questions receiving answers where questions submitted by users with a higher reputation are more likely to be answered and resolved \cite{asaduzzaman2013answering, rahman2015insight, mondal2022reproducibility, calefato2018ask}. We thus consider the users' reputation as a potential factor besides the presence/absence of code snippets and investigate how it affects questions receiving answers.

    The SO data dump only reports the latest reputation scores of the users \cite{mondal2022reproducibility}, which might not be appropriate for our analysis. 
    We thus used a standard equation provided by SO to determine reputation during a question submission \citep{HDR-StackOverflow-2020}. We then divide the users into four categories based on their reputation score \cite{calefato2018ask, mondal2022reproducibility}. They are -- \emph{New} user (score $< 10$), \emph{Low Reputed} user ($10 \le$ score $< 1K$), \emph{Established} user ($1K \le$ score $< 20K$) and \emph{Trusted} user (score $\geq 20K$). 
    Please note that there were only two questions in our dataset submitted by trusted users. We thus discard those two questions and the trusted user category from our analysis.    
    
     $\bullet$ \emph{Question Submission Time:} Several studies suggest that question submission time might affect questions receiving answers \cite{bosu2013building,calefato2018ask, mondal2022reproducibility}. We thus investigate the impact of the question submission time on questions receiving answers. We convert the question submission time to Universal Time Coordinated (UTC) and then divide the submission time into four time frames based on working hours and day. They are - \emph{day}, \emph{night}, \emph{weekday} and \emph{weekend}. 
     

\subsection{Reasons Behind Missing Code Snippets (RQ\textsubscript{2})}
\label{subsec-mothodology:reason-understanding}

We survey software practitioners to understand why the required code snippets might be missed during the submission of a question at SO. 
Such insights might lead to better support for SO users. We follow the guidelines and steps of Kitchenham and Pfleeger~\cite{kitchenham2008personal} on the personal opinion survey. We also consider ethical issues from the established best practices~\cite{groves2011survey, singer2002ethical}. 
For example, we collect participants' consent, assure the confidentiality of their information, and explain the purpose before the survey. We conducted a pilot survey with a couple of practitioners. We collect their feedback on (a) survey duration and (b) clarity and understandability of the questions. We then made minor modifications based on their feedback and prepared the final version of our survey. We excluded these pilot survey responses from the finally analyzed responses.

We recruit 
active users of SO as participants (Table \ref{table:participants-summary}) and select them as follows.


    $\bullet$ \emph{Snowball Approach:}
    We use convenience sampling to bootstrap the snowball \cite{stratton2021population}. First, we contacted a few software developers who were known to us, easily reachable, and working in software companies worldwide. We explain our study goals and share the online survey with them. We then adopt a snowballing method \cite{bi2021accessibility} to disseminate the survey to several of their colleagues with similar experiences.
    
     $\bullet$ \emph{Open Circular:}
     We circulate the survey to specialized Facebook groups. In particular, we target the groups where professional software developers discuss their programming problems. We also use LinkedIn to find potential participants because it is one of the largest professional networks.
     


\subsection{Construction of Prediction Models (RQ\textsubscript{3})}
\label{subsec-mothodology:prediction-models}

\noindent \textbf{Selection of Training \& Test Sets.}
We collect 2,414 samples from two types of questions (1,207 need code snippets + 1,207 do not need code snippets) to construct our prediction models. We select 70\% of the samples (845 that need code + 845 that do not need code) as our training set and the remaining 30\% as the test set.  
In particular, we sort the questions according to their submission date in ascending order and make sure that the training questions are older than the test questions.
The idea was to test our models on an unseen dataset in a realistic setting and keep the train and test datasets separate.

    
    
    
    
    

\begin{table}[htb]
	\centering
	\captionsetup{justification=centering, labelsep=newline}
	\caption{Feature summary (complete list of keywords and key phrases can be found in our online appendix \protect\cite{replicationPackage})}
	\label{table:feature-examples}
	\resizebox{3.4in}{!}{%
        \begin{tabular}{p{4cm}|p{6cm}}
        \hline
        \textbf{Feature} 
        & \multicolumn{1}{c}{\textbf{Examples}} \\ \hline
           Keywords (total \#127)                
           &  access; applic; bug; build; cannot; code; develop; displai; element; error; fail; fine; fix; function; get; help; server; show; tool; try; version; work; wrong  \\ \hline
           Key phrases (POS Patterns) (total \#122: 36 title + 86 body)  
           & $P_{body}$ $\rightarrow$  [adverb] [verb be] [determiner] [noun] 
    
           $P_{body}$ $\rightarrow$  [verb be] [gerund] [determiner] 
    
           $P_{title}$ $\rightarrow$  [wh-adverb] [infinitive to] [verb be] [noun]    \\ \hline
           Sentence structure          
           & I tried to order the rows by code in Ruby on Rails, \textbf{but} this doesn't solve the problem   \\ \hline
           Code elements                
           & \texttt{response.json(); input\_shape; float}   \\ \hline
        \end{tabular}
        }
\vspace{-1mm}
\end{table}

\noindent \textbf{Feature Extraction.}
Table \ref{table:feature-examples} summarizes our extracted features from SO questions.
During dataset construction and manual screening (Section \ref{subsec-mothodology:createDataset}), we notice that several keywords, key phrases (e.g., part of a sentence), and sentence structures might be common in the questions that need code.  
We thus systematically analyze all the training samples and identify keywords, key phrases, and sentence structures from them.

$\bullet$ \emph{Keywords}: 
We first extract the texts from the \emph{title} and \emph{body} of 1,690 questions (845 that need code + 845 that do not need code). Second, we remove both stop words \& code-like elements (i.e., enclosed by \emph{$<$code$>$ ... $</$code$>$} tags) and lowercase the text. Since keywords can be in different forms within the texts, we apply stemmer to reduce the words to their root form. We use Porter's algorithm \cite{porter1980algorithm} for the stemming operation. 
A few keywords could be more frequent in the questions that need code than in those that do not need any code. Therefore, the presence or absence of these keywords in the questions could be leveraged to classify questions. 
We thus attempt to see the prevalence of each word against the entire training dataset of questions that need and do not need any code separately. A word can be a potential keyword candidate if the frequencies \& occurrence possibilities significantly differ between two question categories. We estimate the difference in occurrence possibilities by calculating the frequency ratio.
In particular, we calculate the frequency difference ($d$) and ratio ($r$) for each word as follows.

\smallskip
{\footnotesize{
  $d =
    \begin{cases}
      f_c - f_{nc}\text{, if $f_c \geq f_{nc}$} \\
      f_{nc} - f_c\text{, if $f_{nc} > f_c$} \\
    \end{cases} 
  $
  $r =
    \begin{cases}
      \frac{f_{nc}}{f_c} \times 100\text{, if $f_c \geq f_{nc}$} \\
      \frac{f_c}{f_{nc}} \times 100\text{, if $f_{nc} > f_c$} \\
    \end{cases}  
  $
}
}

\smallskip 
\noindent where $f_c$ \& $f_{nc}$ are the frequencies of a word in the dataset of questions that need and do not need code snippets, respectively.
In particular, we select keywords - (a) which are present at least 50 times more (i.e., $d \geq 50$) in one question category than another and (b) whose occurrence possibility is at least 50\% less (i.e., $r \le 50\%$) in one question category than another. For example, the keyword \emph{error} is available $328$ times in questions that need code snippets (i.e., $f_c (error) = 328$), whereas $f_{nc} (error) = 90$. We thus select \emph{error} since $d = 238  \geq 50$ and $r = 27.4\% \le 50\%$.
In this process, we select a total of $127$ keywords that could classify the questions that need code snippets from those that do not need such code snippets. Table \ref{table:feature-examples} shows a few of our selected keywords.

$\bullet$ \emph{Key phrases (POS Patterns):} 
We investigate whether part-of-speech (POS) patterns could be useful to detect questions with missing code since they were used to detect bug reports with missing information~\cite{chaparro2017detecting}.
First, we extract the texts from the body and title of a question separately.
We split the body text into sentences, whereas the title is considered a single sentence. Then, we apply parts-of-speech (POS) tagging to each sentence using the popular Apache OpenNLP \cite{openNLP} library and collect the POS tags.
Note that we neither remove the stop words nor apply stemming in this case.
According to our investigation, removing stop words can distort the context, and stemming can change the POS tag of words. 
Next, we combine consecutive POS tags 
from each sentence, which is used to detect the POS patterns. 

We then follow a similar approach, like keyword selection, to select the POS-based patterns. We calculate the frequency difference ($d$) and ratio ($r$) of each POS pattern of length 3--6. We select POS patterns from body text when $d \geq 50$ and $r \le 50\%$. However, we choose $d \geq 20$ for selecting title patterns since the titles contain limited text. Finally, we get $36$ patterns for the title and $86$ for the body. We did not choose any patterns of length two (i.e., bigram) because they are close to keywords (i.e., unigram). However, we did not find any patterns of length more than six that follow our selection constraints. Table \ref{table:feature-examples} shows a few of our selected POS patterns.

$\bullet$ \emph{Sentence structure:}
Our manual investigation reveals a complex sentence structure where two clauses are connected with several conjunctions -- \emph{but, however, except, while, when, because}. These sentences usually discuss the unexpected behaviors of code, coding errors, users' efforts, and inabilities to solve problems. Table \ref{table:feature-examples} shows an example sentence.
Such sentences are available in both question types. However, they are more frequent in the question descriptions that need code snippets. We thus count these sentences from the body text as a feature to classify the question types. 

$\bullet$ \emph{Code elements:}
We also see that a question with more code-like elements is more likely to discuss code-related problems and vice versa. We thus mine code-like elements (e.g., method name, identifier) from the contents of each question. In SO, code elements are often placed inside the \emph{$<$code$>$...$<$/code$>$} tag. 
We thus mine such tags carefully and capture the occurrences of
code elements from each question. 

\noindent \textbf{Model Selection.}
Existing studies
show that supervised ML algorithms (e.g., Random Forest) can perform equally well or even better than deep learning techniques 
when a small data set is used \cite{fu2017easy, beyer2018automatically}.
We thus choose six popular supervised ML techniques with different learning strategies. They are -- (a) Random Forest (RF)~\cite{randomForest}, (b) eXtreme Gradient Boosting (XGBoost)~\cite{chen2015xgboost}, (c) Artificial Neural Network (ANN)~\cite{ann}, (d) Gaussian Naive Bayes (NB)~\cite{webb2010naive, rish2001empirical},  (e) K-Nearest Neighbors (KNN)~\cite{knn}, 
and (f) Support Vector Machines (SVM)~\cite{svm}. In particular, we choose these ML algorithms
because they were widely used in the relevant studies \cite{TUT-Saha-2013,UAC-Ponzanelli-2014,AII-Rahman-2015, beyer2018automatically, mondal2023automatic, mondal2023subjectivity}.  

\noindent \textbf{Performance Analysis.}
We use the default 
implementation of each algorithm from the \emph{Scikit-learn} \cite{SML-Pedregosa-2011} library to train our models.
We then evaluate the performance of our prediction models using four performance metrics that are widely used in the literature \cite{mondal2023automatic, mondal2023subjectivity, beyer2018automatically}. They are -- (1) \emph{precision} that measures the ratio of correctly classified questions into a class (need code/do not need code) with respect to all questions classified into that class, (2) \emph{recall} that measures the ratio of correctly classified questions with respect to the actually observed questions as true instances, (3) \emph{F1-score} that offers the harmonic mean of precision and recall, and (4) \emph{classification accuracy} that measures the ratio of correctly classified questions into true \& false classes with respect to all classified questions. 

\section{Study Findings}
\label{sec:study-findings}

\subsection{Effect Analysis of Missing Code Snippets (RQ1)} 
\label{subsec-findings:impactAnalysis}



In this section,  we analyze the correlation between three types of questions -- MICO, COAC \& CODS and corresponding answer meta-data. In particular, we divide \textbf{RQ\textsubscript{1}} into four sub-questions and answer them with detailed statistics.








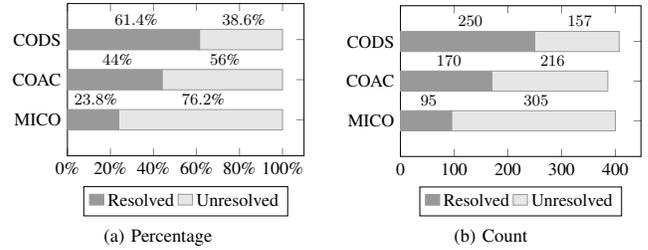
\begin{figure}[!htp]
    \centering
    \subfloat[Percentage]{
    \label{fig:accepted-answer-ratio}
    \resizebox{1.65in}{!}{
    \begin{tikzpicture}
    \begin{axis}[
        xmin=0,
        width=2.45in,
        height=1.8in,
        enlarge y limits=0.45,
        bar width=0.40cm,
        symbolic y coords={MICO, COAC, CODS},
        ytick=data,
        xtick={0,20,...,100},
        xticklabels={0\%,20\%,40\%,60\%,80\%,100\%,},
        xbar stacked,
        legend style={at={(0.5,-0.20)},
          anchor=north,legend columns=-1},
        ytick=data,
        every node near coord/.style={
            font=\small,
            black!100,
        },
        nodes near coords align=center,
        nodes near coords style={yshift= 0.4cm, rotate=0},
        xbar stacked, nodes near coords={\hspace{0cm} \pgfmathprintnumber[precision=1]{\pgfplotspointmeta}\%}
        ]
    \addplot+[xbar, fill=gray!80,draw=gray!80] plot coordinates {(23.8,MICO) (44,COAC) (61.4,CODS)};
    \addplot+[xbar, fill=gray!20,draw=gray!80] plot coordinates {(76.2,MICO) (56,COAC) (38.6,CODS)};
    
    \legend{Resolved, Unresolved}
    
    \end{axis}
    \end{tikzpicture}
    }
    }
    \subfloat[Count]{
    \label{fig:accepted-answer-count}
    \resizebox{1.64in}{!}{
    \begin{tikzpicture}
    \begin{axis}[
        xmin=0,
        width=2.5in,
        height=1.8in,
        enlarge y limits=0.45,
        bar width=0.40cm,
        symbolic y coords={MICO, COAC, CODS},
        ytick=data,
        xtick={0,100,...,400},
        xticklabels={0,100,200,300,400},
        xbar stacked,
        legend style={at={(0.5,-0.20)},
          anchor=north,legend columns=-1},
        ytick=data,
        every node near coord/.style={
            font=\small,
            black!100,
        },
        nodes near coords align=center,
        nodes near coords style={yshift= 0.4cm, rotate=0},
        xbar stacked, nodes near coords={\hspace{0cm} \pgfmathprintnumber[precision=0]{\pgfplotspointmeta}}
        ]
    \addplot+[xbar, fill=gray!80,draw=gray!80] plot coordinates {(95,MICO) (170,COAC) (250,CODS)};
    \addplot+[xbar, fill=gray!20,draw=gray!80] plot coordinates {(305,MICO) (216,COAC) (157,CODS)};
    
    \legend{Resolved, Unresolved}
    
    \end{axis}
    \end{tikzpicture}
    }
    }
\caption{Percentage and count of resolved \& unresolved questions}
\label{fig:accepted-answer-ratio-count}
\vspace{-5mm}
\end{figure}


\noindent \textbf{RQ\textsubscript{1}(a): Does the inclusion of required code snippets in Stack Overflow questions encourage acceptable answers?}
Fig. \ref{fig:accepted-answer-ratio-count} shows the percentage (Fig. \ref{fig:accepted-answer-ratio}) and count (Fig. \ref{fig:accepted-answer-count}) of \emph{resolved} (i.e., received acceptable answers) and \emph{unresolved} (i.e., did not receive acceptable answers) questions from each category. We see that the presence of code snippets in the questions leads to more acceptable answers. 
We find that 61.4\% (250 out of 407) of questions where required code snippets were included during submission received acceptable answers (i.e., appropriate solutions). 
Questions whose code snippets were submitted after the requests received 184 (out of 400) acceptable answers. Among them, 14 answers were submitted before adding the code snippets, which were discarded from our analysis. 
Nevertheless, 44\% (170 out of 386) of questions whose code snippets were submitted after the requests received accepted answers. On the other hand, only 23.8\% (95 out of 400) of questions received acceptable answers when they missed the required code snippets and did not include them even after the requests. Thus, the inclusion of required code snippets during the submission of a question increases its chance of getting an acceptable solution almost three times. Moreover, the chance of getting a solution is double if the code is added after a request from fellow SO users.

   
    
    
    

We also investigate how receiving of acceptable answers depends on the inclusion of required code snippets.
In particular, we identify two categorical variables -  \emph{question category} and \emph{presence of acceptable answer} and their frequencies from Fig. \ref{fig:accepted-answer-count}. We use \emph{Chi-Squared} \cite{mchugh2013chi} statistical test to measure the independence between these two variables.
We find statistically significant \emph{p-value} ($p-value \approx 0 < 0.05$) from our analysis. Thus, there is a strong positive correlation between the inclusion of required code snippets in the questions and their chance of getting acceptable answers.

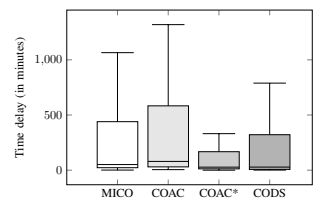
\begin{figure}[!htb]
    \centering
        \resizebox{1.6in}{!}{%
        \begin{tikzpicture}
        \begin{axis}[xmin=0,width=3.4in,height=2.6in,
          ylabel = {Time delay (in minutes)},
          xtick = {1, 2, 3, 4},
          xticklabels = {MICO, COAC, COAC*, CODS},
          ]
        \addplot+[boxplot, 
                  /pgfplots/boxplot/hide outliers, 
                  boxplot/draw direction = y,
                  mark = *,
                  boxplot,
                  fill,fill opacity=0.0,
                  black]
        table[row sep=\\,y index=0] {
        data\\
        83\\
        496\\
        7\\
        542\\
        169\\
        849\\
        29\\
        30\\
        47\\
        41\\
        20\\
        29\\
        3\\
        2\\
        96\\
        568\\
        3358\\
        1072\\
        45\\
        217144\\
        4535\\
        9\\
        106\\
        123941\\
        80\\
        3\\
        214\\
        1235\\
        10\\
        418\\
        217\\
        42\\
        51\\
        143\\
        11459\\
        104\\
        446\\
        41\\
        191\\
        155\\
        10416\\
        40\\
        1645\\
        42\\
        6\\
        2369\\
        10\\
        61\\
        1065\\
        22\\
        159\\
        17\\
        663\\
        29\\
        711\\
        1650\\
        74\\
        30\\
        1\\
        24\\
        10\\
        9825\\
        41\\
        71\\
        191\\
        48\\
        62\\
        5\\
        232\\
        4\\
        5\\
        17\\
        28\\
        14\\
        35\\
        23\\
        35\\
        106200\\
        433\\
        7\\
        576472\\
        185\\
        4\\
        26\\
        4\\
        386\\
        4155\\
        38\\
        39\\
        5\\
        461\\
        15\\
        13\\
        28\\
        73\\
        };
        
        \addplot+[boxplot, 
                  /pgfplots/boxplot/hide outliers, 
                  boxplot/draw direction = y,
                  mark = *,
                  boxplot,
                  fill,fill opacity=0.1,
                  black]
        table[row sep=\\,y index=0] {
        data\\
        248\\
        79\\
        125\\
        5209\\
        42\\
        80\\
        529\\
        6028\\
        1140\\
        30\\
        54\\
        170\\
        3185\\
        317\\
        13\\
        68\\
        33\\
        5\\
        19\\
        1997338\\
        71\\
        22\\
        57\\
        1468\\
        766\\
        17\\
        25\\
        995\\
        29\\
        3436\\
        248\\
        72\\
        28\\
        22\\
        3691\\
        2758\\
        28\\
        56\\
        40\\
        30\\
        14\\
        31\\
        47\\
        3522\\
        159\\
        47\\
        14\\
        105\\
        4827\\
        17\\
        84\\
        81\\
        15\\
        728\\
        44\\
        11\\
        59\\
        16\\
        35\\
        341\\
        28\\
        39\\
        94\\
        97\\
        1319\\
        51\\
        10305\\
        75\\
        2336\\
        92\\
        747\\
        22\\
        3274\\
        10\\
        335\\
        238\\
        22\\
        185\\
        601\\
        75\\
        126\\
        227\\
        220\\
        878\\
        142\\
        10\\
        1033\\
        10\\
        305\\
        45\\
        183\\
        4752\\
        16\\
        86\\
        38\\
        730\\
        15\\
        1022\\
        18\\
        372\\
        37\\
        4660\\
        3116\\
        1090\\
        71\\
        720\\
        30\\
        43\\
        186\\
        1551\\
        48\\
        1184\\
        180\\
        164\\
        40\\
        3358\\
        27\\
        1023\\
        1794\\
        125\\
        62\\
        24\\
        35\\
        55\\
        32\\
        201\\
        20\\
        43\\
        25\\
        76\\
        126\\
        1835\\
        38\\
        152\\
        26\\
        16\\
        17\\
        117\\
        12\\
        1662\\
        195\\
        37\\
        54\\
        2040\\
        23\\
        107\\
        9\\
        910\\
        87\\
        465\\
        15\\
        101\\
        9\\
        30\\
        24\\
        26\\
        39\\
        18\\
        96\\
        101\\
        50\\
        1475\\
        1427\\
        1130\\
        66\\
        5732\\
        21\\
        3381\\
        35\\
        89\\
        };
        
        \addplot+[boxplot, 
                  /pgfplots/boxplot/hide outliers, 
                  boxplot/draw direction = y,
                  mark = *,
                  boxplot,
                  fill,fill opacity=0.2,
                  black]
        table[row sep=\\,y index=0] {
        data\\
        232\\
        44\\
        110\\
        2528\\
        19\\
        37\\
        26\\
        464\\
        1102\\
        22\\
        11\\
        126\\
        2456\\
        171\\
        6\\
        18\\
        30\\
        2\\
        16\\
        1997305\\
        17\\
        16\\
        22\\
        1412\\
        649\\
        3\\
        16\\
        987\\
        2\\
        1808\\
        242\\
        67\\
        22\\
        17\\
        3628\\
        31\\
        23\\
        7\\
        25\\
        22\\
        8\\
        23\\
        1\\
        3508\\
        107\\
        27\\
        8\\
        73\\
        4238\\
        13\\
        52\\
        31\\
        5\\
        691\\
        21\\
        6\\
        20\\
        0\\
        15\\
        37\\
        18\\
        5\\
        13\\
        92\\
        1229\\
        33\\
        3703\\
        71\\
        1372\\
        54\\
        21\\
        11\\
        3259\\
        1\\
        291\\
        15\\
        8\\
        61\\
        595\\
        60\\
        86\\
        212\\
        10\\
        830\\
        92\\
        3\\
        868\\
        6\\
        5\\
        6\\
        118\\
        4743\\
        4\\
        61\\
        18\\
        520\\
        13\\
        331\\
        5\\
        8\\
        20\\
        95\\
        3051\\
        13\\
        4\\
        331\\
        12\\
        7\\
        146\\
        1493\\
        41\\
        112\\
        157\\
        121\\
        3\\
        1527\\
        23\\
        958\\
        691\\
        89\\
        27\\
        12\\
        26\\
        20\\
        15\\
        195\\
        4\\
        19\\
        7\\
        26\\
        38\\
        1795\\
        31\\
        73\\
        18\\
        1\\
        10\\
        50\\
        5\\
        1580\\
        34\\
        20\\
        28\\
        102\\
        2\\
        81\\
        4\\
        850\\
        21\\
        187\\
        1\\
        77\\
        2\\
        11\\
        11\\
        3\\
        0\\
        4\\
        75\\
        66\\
        19\\
        1444\\
        1342\\
        277\\
        39\\
        12\\
        7\\
        1786\\
        25\\
        48\\
        };
        
        \addplot+[boxplot, 
                  /pgfplots/boxplot/hide outliers, 
                  boxplot/draw direction = y,
                  mark = *,
                  boxplot,
                  fill,fill opacity=0.3,
                  black]
        table[row sep=\\,y index=0] {
        data\\
        9\\
        250\\
        2\\
        7\\
        18\\
        5\\
        326686\\
        71\\
        1584\\
        2\\
        12\\
        20\\
        2\\
        60\\
        59\\
        49\\
        5\\
        65\\
        12\\
        1075885\\
        3\\
        5\\
        1035492\\
        16\\
        288534\\
        264\\
        5\\
        663\\
        10\\
        11\\
        85\\
        559\\
        10\\
        17\\
        26\\
        9\\
        42\\
        550\\
        5\\
        380\\
        7\\
        7\\
        27\\
        12\\
        11\\
        8\\
        118\\
        95\\
        559\\
        317\\
        41\\
        3\\
        5\\
        445\\
        963\\
        3074\\
        1422\\
        182\\
        3475\\
        4009\\
        449\\
        21\\
        3367\\
        323\\
        390\\
        3926\\
        1876\\
        2111\\
        3\\
        427\\
        253\\
        2087\\
        19\\
        952\\
        4\\
        1047\\
        4604127\\
        2\\
        853\\
        1469\\
        334\\
        5\\
        20\\
        2\\
        4\\
        216\\
        150\\
        158067\\
        108\\
        51\\
        533\\
        10\\
        147\\
        5\\
        35\\
        13\\
        1\\
        1\\
        11\\
        71\\
        40\\
        691\\
        5\\
        1740\\
        1041\\
        66\\
        202\\
        804351\\
        9\\
        35\\
        5\\
        92\\
        607533\\
        21\\
        647\\
        148\\
        231\\
        19\\
        113\\
        4\\
        17\\
        22\\
        7\\
        892\\
        784\\
        8\\
        282\\
        11\\
        621\\
        7\\
        2\\
        772\\
        12\\
        9\\
        915\\
        21\\
        1040\\
        621\\
        22\\
        18\\
        3\\
        8\\
        151\\
        6\\
        21\\
        10\\
        4\\
        119\\
        6\\
        186\\
        259\\
        283\\
        1299\\
        1333\\
        27\\
        8\\
        956\\
        104\\
        18\\
        41\\
        7\\
        72\\
        198\\
        6\\
        76\\
        4\\
        26\\
        20\\
        4\\
        129\\
        1293\\
        1\\
        21\\
        39\\
        270\\
        13\\
        67\\
        4\\
        675488\\
        6\\
        1\\
        5\\
        4\\
        64\\
        4\\
        109\\
        47\\
        31\\
        5\\
        76\\
        18\\
        12\\
        3\\
        19\\
        4\\
        11\\
        16\\
        16\\
        17\\
        28\\
        243\\
        5\\
        633\\
        4\\
        422\\
        4\\
        5\\
        10\\
        3091\\
        789\\
        2640\\
        2393\\
        1287\\
        126\\
        1841\\
        3108\\
        3\\
        84\\
        538\\
        1907\\
        46\\
        6\\
        35\\
        50\\
        100\\
        11\\
        1287\\
        38\\
        6\\
        44\\
        39\\
        5\\
        18\\
        1\\
        28\\
        1\\
        4\\
        19\\
        3\\
        4\\
        9\\
        10\\
        6\\
        4\\
        8\\
        62\\
        5\\
        10\\
        2\\
        7\\
        };
        \end{axis}
        \end{tikzpicture}
        }
    \caption{Time delay of receiving acceptable answers.}
    \label{fig:delay-receiving-acc-ans}
\vspace{-2mm}
\end{figure}

\noindent \textbf{RQ\textsubscript{1}(b): Does the inclusion of required code snippets in Stack Overflow questions reduce the time delay in getting acceptable answers?}
According to \textbf{RQ\textsubscript{1}a}, there exists a strong correlation between the inclusion of required code snippets in the questions and their chance of getting an acceptable answer. 
This section investigates whether the inclusion of required code snippets influences the delay in getting an acceptable answer. First, we calculate the gap between the submission time of questions and that of the accepted answers. For the COAC category, we also determine the delay in receiving
the accepted answers once the required code snippets are included in the questions. Hereby, we name it \textbf{COAC*} to avoid confusion. Then, we contrast the question categories using such delays. 


    

    
    

Fig. \ref{fig:delay-receiving-acc-ans} shows the box plots of delays in getting acceptable answers for three question types.
We see that the median delay in getting an accepted answer is only 19 minutes for the questions of the CODS category. On the contrary, the delay is more than double (i.e., 41 minutes) for the MICO category. The delay is 56 minutes for COAC.
However, the median delay reduces to only 21 minutes if we calculate the delay between including the required code snippets in the questions and getting acceptable answers.
Such findings clearly suggest that including required code snippets into questions significantly reduces the delay in getting the accepted answers.

\begin{table}[!htp]
\centering
    \caption{Statistical tests summary of the time delay of receiving accepted answer}
	\label{table:statistical-tests-summary-delay}
	\resizebox{3in}{!}{%
    \begin{tabular}{l|l|l} \hline
   \multicolumn{1}{c|}{\textbf{Group}} & \multicolumn{1}{c|}{\textbf{p-value}} & \multicolumn{1}{c}{\textbf{Effect Size}} \\  \hline
   
    \textbf{$G_1$: MICO \& COAC}     & 0.04 $< 0.05$ (significant)   & 0.16 (small) \\  \hline
    \textbf{$G_2$: COAC \& CODS}     & 0 $< 0.05$ (significant)      & 0.39 (medium)  \\  \hline
    \textbf{$G_3$: MICO \& CODS}     & 0 $< 0.05$ (significant)      &  0.22 (small) \\  \hline
    \textbf{$G_4$: MICO \& COAC*}    & 0 $< 0.05$ (significant)      & 0.30 (small)  \\  \hline
    \textbf{$G_5$: COAC* \& CODS}    & 0.73 $> 0.05$ (insignificant) &  0.02 (negligible) \\  \hline

    \end{tabular}
    }
    \vspace{-1mm}
\end{table}


We also use the \emph{Mann-Whitney-Wilcoxon} test to check whether the difference in delays is statistically significant. We also measure their effect size using \emph{Cliff's Delta} test. 
Table \ref{table:statistical-tests-summary-delay} summarizes our test results.
We see that all the differences are statistically significant with small to medium effect sizes except $G_5$. Given the evidence above, the inclusion of required code snippets significantly reduces the delay in getting acceptable answers.

\begin{figure}[!htp]

        \centering
        \resizebox{2in}{!}{
        \begin{tikzpicture}
        \begin{axis}[
            xmin=0,
            width=2.5in,
            height=1.9in,
            enlarge y limits=0.45,
            bar width=0.40cm,
            symbolic y coords={MICO, COAC, CODS},
            ytick=data,
            xtick={0,20,...,100},
            xticklabels={0\%,20\%,40\%,60\%,80\%,100\%},
            xbar stacked,
            legend style={
              anchor=north,legend columns=1},
              legend pos = outer north east,
            ytick=data,
            every node near coord/.style={
                font=\small,
                black!100,
            },
            nodes near coords align=center,
            nodes near coords style={yshift= 0.4cm, rotate=0},
            xbar stacked, nodes near coords={\hspace{0cm} \pgfmathprintnumber[precision=1]{\pgfplotspointmeta}\%}
            ]
        \addplot+[xbar, fill=gray!80,draw=gray!80] plot coordinates {(72,MICO) (80,COAC) (91.2,CODS)};
        \addplot+[xbar, fill=gray!20,draw=gray!80] plot coordinates {(28,MICO) (20,COAC) (8.8,CODS)};
        
        \legend{Answered, Unanswered}
        
        \end{axis}
        \end{tikzpicture}
        }
        \caption{Percentage of unanswered questions}
        \label{fig:unanswered-question-ratio}
\vspace{-2mm}
\end{figure}
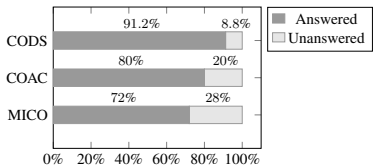

\noindent \textbf{RQ\textsubscript{1}(c): Does the inclusion of required code snippets in Stack Overflow questions encourage answers?}
According to the previous sections, the inclusion of required code snippets in the questions might encourage quick and high-quality responses from the users. This section further investigates whether such inclusion of required code also encourages answers at SO.
Fig. \ref{fig:unanswered-question-ratio} summarizes our findings. We see that 91.2\% (371 out of 407) of the questions that contain code segments during their submission (a.k.a., CODS) receive one or more answers. The percentage is 80\% (320 out of 400) for COAC. However, such a percentage reduces to 72\% (288 out of 400) for the questions that miss required code altogether (a.k.a., MICO). Thus, the inclusion of required code snippets in the questions increases their chance of receiving answers. We also use the \emph{Chi-Squared} test to measure the independence of these two categorical variables -- \emph{question category} and \emph{presence of answers} statistically. We find statistical significance \emph{p-value} ($p-value = 0 < 0.05$), which indicates a strong positive correlation between the inclusion of required code snippets in questions and their chance of getting answers.

\begin{table}[htb]
\centering
    \caption{Question category vs. (a) presence of accepted answer (b) time delay of receiving accepted answer and (c) percentage of unanswered questions according to the reputation of the question submitter}
	\label{table:compounding-variable-reputation-accepted-answer-time-delay-no-of-answers}
        \subfloat[Percentage of accepted answer]{
        \label{table:compounding-variable-reputation-accepted-answer}
        \resizebox{2.5in}{!}{%
        \begin{tabular}{l|c|c|c}
        \hline
        \multirow{2}{*}{\textbf{Category}} & \multicolumn{3}{c}{\textbf{User Category}} \\ 
                                           & \textbf{New}      & \textbf{Low Reputed} & \textbf{Established}  \\ \hline
        \textbf{CODS}                      &  39/81 (48.1\%)   & 160/260 (61.5\%)    & 50/64 (78.1\%)       \\ \hline
        \textbf{COAC}                      &  50/130 (38.5\%)  & 121/248 (48.8\%)    & 13/22 (59.1\%)       \\ \hline
        \textbf{MICO}                      &  29/195 (14.9\%)  & 60/188 (31.9\%)      & 6/17 (35.3\%)         \\ \hline
        \end{tabular}
        }
        }
	\vspace{-2mm}
 
	\subfloat[Time Delay]{
	\label{table:compounding-variable-reputation-delay}
	\resizebox{2.4in}{!}{%
        \begin{tabular}{l|c|c|c|c|c|c}
        \hline
        \multirow{3}{*}{\textbf{Category}} & \multicolumn{6}{c}{\textbf{User Category}}                                                                                                                                                                                                                                                               \\ 
                                                & \multicolumn{2}{c|}{\textbf{New}} & \multicolumn{2}{c|}{\textbf{Low Reputed}} & \multicolumn{2}{c}{\textbf{Established}}    \\ 
                                                & \textbf{Med}  & \textbf{Avg} & \textbf{Med} & \textbf{Avg} & \textbf{Med} & \textbf{Avg}\\ \hline
                                                
        \textbf{CODS}         & 11.0 &	38.3 &	17.5 &	117.6 &	27.0 &	119.0   \\ \hline
        \textbf{COAC}         & 36.0 &	243.2 &	58.0 &	165.0 &	71.0 &	119.7   \\ \hline
        \textbf{COAC*}        & 22.5 &	228.9 &	22.0 &	55.9 &	6.0 &	46.6            \\ \hline
        \textbf{MICO}         & 29.0 &	86.2 &	44.5 &	179.6 &	35.0 &	134.8    \\ \hline
        \end{tabular}
        }
        }
        \vspace{-2mm}

        \subfloat[Percentage of unanswered questions]{
        \label{table:compounding-variable-reputation-no-of-answers}
        \resizebox{2.5in}{!}{%
        \begin{tabular}{l|c|c|c}
        \hline
        \multirow{2}{*}{\textbf{Category}} & \multicolumn{3}{c}{\textbf{User Category}} \\ 
                                           & \textbf{New}      & \textbf{Low Reputed} & \textbf{Established}  \\ \hline
        \textbf{CODS}                      &  9/81 (11.1\%)   & 24/260 (9.2\%)    & 3/64 (4.7\%)       \\ \hline
        \textbf{COAC}                      &  29/130 (22.3\%)  & 45/248 (18.1\%)    & 6/22 (27.3\%)       \\ \hline
        \textbf{MICO}                      &  48/195 (24.6\%)  & 60/188 (31.9\%)      & 4/17 (23.5\%)         \\ \hline 
        \end{tabular}
        }
    }
\vspace{-7mm}
\end{table}

\noindent \textbf{RQ\textsubscript{1}(d): What factors affect questions receiving answers besides including required code?}
Previous sections show that including required code snippets
in the questions encourages quick and high-quality answers. In this section, we investigate whether user reputation and question submission time affect questions receiving answers.

\noindent\textbf{Reputation.}
Table \ref{table:compounding-variable-reputation-accepted-answer} shows how user reputation affects questions receiving accepted answers besides the presence or absence of required code snippets. We see that questions submitted by users with a higher reputation have a higher chance of receiving accepted answers than those submitted by new users. For example, questions from the CODS category receive 78.1\% of accepted answers if they were submitted by established users, as opposed to 48.1\% if new users submitted them. We see similar results for the COAC and MICO categories. However, regardless of reputation, the chance of receiving accepted answers to questions that included required code snippets during submission or after the request is two to three times higher than those missed required code snippets. Such findings indicate the importance of adding required code snippets. Even new users receive 48.1\% acceptable answers to their questions when they add required code snippets during submission, as opposed to 20.5\% when they miss it.

Next, we investigate whether reputation affects the delay in receiving accepted answers. Table \ref{table:compounding-variable-reputation-delay} shows the median and average time delay (in minutes) of receiving accepted answers for each user category. In some cases, questions submitted by users with a lower reputation receive accepted answers with a minimum delay. For example, the median time delay of receiving accepted answers to questions that included code during submission is $17.5$ minutes when submitted by low-reputed users but $27$ minutes for established users. 
The opposite scenario was also seen for the MICO category, where the median delay in receiving accepted answers to questions submitted by low-reputed users is higher than those submitted by established users. Surprisingly, the delay is comparatively lower in receiving accepted answers to questions submitted by new users.
However, more interestingly, the time delay is controlled by the presence or absence of required code snippets. The time delay in receiving accepted answers to questions that included required code snippets is significantly lower than those that missed it. For example, for the new user category, the median time delay in receiving accepted answers is $11$ minutes for questions that included required code snippets during submission. On the contrary, such median time delay is as high as $29$ minutes for the questions that miss code snippets. 

Finally, we attempt to see whether reputation influences questions getting answers. As shown in Table \ref{table:compounding-variable-reputation-no-of-answers}, for the CODS category, questions submitted by users with a higher reputation are more likely to be answered. For example, the percentage of unanswered questions is only 4.7\% that were submitted by established users as opposed to 11.1\% submitted by new users. We see mixed results for the COAC and MICO categories. However, the rate of questions remaining unanswered is two to five times more that missed required code snippets than those included code snippets during submission.


    


\noindent\textbf{Question Submission Time.} Table \ref{table:compounding-variable-question-submission-time-accepted-answer} shows the rate of receiving accepted answers for different time slots. We see that questions submitted during the day and on weekdays have a slightly higher accepted answer ratio than those submitted on the night and weekend. For example, the percentage of the accepted answers for questions of the COAC category is about 50\% when submitted on weekdays, which is about 37\% if submitted on weekends. However, regardless of the submission time, the chance of receiving accepted answers to questions that included code snippets during submission or after the request is significantly higher than those missed required code snippets.

\begin{table}[htb]
\centering
    \caption{Question category vs. (a) presence of accepted answer (b) time delay of receiving accepted answer and (c) percentage of unanswered questions according to the question submission time}
	\label{table:compounding-variable-question-submission-accepted-answer-time-delay-no-of-answers}

        \subfloat[Percentage of accepted answer]{
        \label{table:compounding-variable-question-submission-time-accepted-answer}
        \resizebox{3in}{!}{%
        \begin{tabular}{l|c|c|c|c}
        \hline
        \multirow{2}{*}{\textbf{Category}} & \multicolumn{2}{c|}{\textbf{Working Hour}}  & \multicolumn{2}{c}{\textbf{Working Day}} \\ 
                    & \textbf{Day}      & \textbf{Night} & \textbf{Weekday} & \textbf{Weekend}  \\ \hline
        \textbf{CODS}   & 153/239 (64.0\%) & 97/168 (57.7\%)& 226/265 (61.9\%) & 24/42 (57.1\%)    \\ \hline
        \textbf{COAC}   & 148/319 (46.4\%) & 36/81 (44.4\%)& 149/305 (48.9\%) & 35/95 (36.8\%)    \\ \hline
        \textbf{MICO}   & 60/293 (20.5\%)   & 35/107 (32.7\%) & 71/291 (24.4\%)  & 24/109 (22.0\%)   \\ \hline
        \end{tabular}
        }
        }
        \vspace{-2mm}
        
	\subfloat[Time Delay]{
	\label{table:compounding-variable-question-submission-time-delay}
	\resizebox{3in}{!}{%
        \begin{tabular}{l|c|c|c|c|c|c|c|c}
        \hline
        \multirow{3}{*}{\textbf{Repro. Status}} & \multicolumn{4}{c|}{\textbf{Working Hour}} & \multicolumn{4}{c}{\textbf{Working Day}}                                                                                                                                                                                                                                                               \\ 
                                                & \multicolumn{2}{c|}{\textbf{Day}} & \multicolumn{2}{c|}{\textbf{Night}} & \multicolumn{2}{c|}{\textbf{Weekday}}  & \multicolumn{2}{c}{\textbf{Weekend}}                                        \\ 
                                                & \textbf{Med}  & \textbf{Avg} & \textbf{Med} & \textbf{Avg} & \textbf{Med} & \textbf{Avg} & \textbf{Med} & \textbf{Avg} \\ \hline
                                                
        \textbf{CODS}          & 19.0 &	100.3 &	17.5 &	79.8 &	18.0 &	87.6 &	10.0 &	162.2   \\ \hline
        \textbf{COAC}          & 45.0 &	102.1 &	76.0 &	366.1 &	47.0 &	110.2 &	43.0 &	112.9  \\ \hline
        \textbf{COAC*}         & 20.0 &	46.0 &	31.0 &	118.5 &	22.0 &	47.6 &	18.5 &	56.0  \\ \hline
        \textbf{MICO}          & 41.0 &	170.2 &	39.0 &	60.3 &	35.0 &	102.8 &	75.5 &	201.8 \\ \hline
        \end{tabular}
        }
        }
        \vspace{-2mm}

        \subfloat[Percentage of unanswered questions]{
        \label{table:compounding-variable-question-submission-time-no-of-answers}
        \resizebox{3in}{!}{%
        \begin{tabular}{l|c|c|c|c}
        \hline
        \multirow{2}{*}{\textbf{Category}} & \multicolumn{2}{c|}{\textbf{Working Hour}}  & \multicolumn{2}{c}{\textbf{Working Day}} \\ 
                    & \textbf{Day}      & \textbf{Night} & \textbf{Weekday} & \textbf{Weekend}  \\ \hline
        \textbf{CODS}   & 90/293 (8.8\%)  & 22/107 (8.9\%) & 77/291 (8.8\%)  & 35/109 (9.5\%)   \\ \hline
        \textbf{COAC}   & 64/319 (20.1\%) & 16/81 (19.8\%)  & 65/305 (21.3\%) & 15/95 (15.8\%)    \\ \hline
        \textbf{MICO}   & 21/239 (30.7\%) & 15/168 (20.6\%) & 32/265 (26.5\%) & 4/42 (32.1\%)   \\ \hline
        \end{tabular}
        }
        }
\vspace{-6mm}
\end{table}

We then investigate whether question submission time (a) impacts the delay in receiving accepted answers and (b) encourages answers. As shown in Table \ref{table:compounding-variable-question-submission-time-delay} \& Table \ref{table:compounding-variable-question-submission-time-no-of-answers}, question submission time has little impact on the delay in receiving accepted answers and encouraging answers. However, we see that including required code snippets consistently reduces the time delay in receiving accepted answers and increases the chance of getting answers.

Our findings might align with the common understanding of the community regarding questions and their code requirements at SO.
However, our in-depth analysis shows concrete empirical evidence of how the answers get affected when the questions miss the required code snippets at SO.

\begin{tcolorbox}[colframe=black!50, colback=white,left=0pt,right=1pt,top=1pt,bottom=1pt, arc=0.5pt]

    \emph{\textbf{Summary of RQ1:}} Questions that include required code snippets during their submission have a three times higher chance of getting acceptable answers than those that miss code snippets. Moreover, the delay is more than double in getting acceptable answers, and the percentage of unanswered questions is significantly high when the questions miss their required code snippets. Other factors (e.g., reputation) might impact questions receiving answers. However, including required code snippets consistently increases the chance of getting prompt and acceptable answers.

\end{tcolorbox}
\vspace{-2mm}

\begin{table}[!htp]
\centering
    \caption{Reasons behind missing the code snippets with Stack Overflow questions whenever required}
	\label{table:reasons-missing-code}
	\resizebox{3.4in}{!}{%
    \begin{tabular}{p{6.5cm}|c|c|c} \hline
    
    \multicolumn{1}{c|}{\textbf{Reasons}} & \textbf{Agree} & \textbf{Neutral} & \textbf{Disagree} \\  \hline
    
    (a) Users are not aware of whether their questions need any code snippets                                       & 59.4\% & 26.6\% & 14.0\% \\ \hline
    (b) Users do not have example code snippets ready when submitting their questions                                      & 29.7\% & 29.7\% & 40.6\% \\ \hline
    (c) There are restrictions from employers to upload code snippets in a public Q\&A site like Stack Overflow & 45.3\% & 31.3\% & 23.4\% \\ \hline
    (d) Code snippets might disclose confidential information                                                   & 45.3\% & 25.0\% & 29.7\% \\ \hline
    (e) Users are busy preparing appropriate example code snippets to support their problem description           & 42.2\% & 31.2\% & 26.6\% \\ \hline
    (f) Users cannot understand which portion of code they should include to support their problem description    & 51.6\% & 29.7\% & 18.7\% \\ \hline

    \end{tabular}
    }
\vspace{-5mm}
\end{table}

\begin{table}[!htp]
\centering
    \caption{Other reasons (excluding the reasons in Table \ref{table:reasons-missing-code}) behind missing the code snippets}
	\label{table:other-reasons-missing-code}
	\resizebox{3.4in}{!}{%
        \begin{tabular}{p{8cm} | c } \hline
        \multicolumn{1}{c|}{\textbf{Reasons}} & \textbf{Count} \\  \hline

        Laziness or negligence in including code snippets & 9 (30.0\%) \\


        Unaware of the negative impact of missing necessary code snippets & 7 (23.3\%) \\

        Creating a representative code snippet from a large source code is challenging & 6 (20.0\%) \\

        The problem could be understood without code snippets (i.e., common problem) & 4 (13.3\%) \\  

        Unaware of SO question submission guidelines & 4 (13.3\%) \\

        Miscellaneous (e.g., Code integration from multiple files/projects is challenging) & 7 (23.3\%) \\ \hline

        \end{tabular}
    }
\vspace{-3mm}

\end{table}





\subsection{Perceived Reasons Behind Missing the Required Code Snippets (RQ2)}
\label{subsec-findings:reason-understanding}

In our survey, we offer six potential reasons with three options (\emph{agree, neutral, disagree}) to the participants. Before our pilot study, we interviewed five SO users to identify these reasons. The estimated time to complete the survey was ten minutes. We received a total of 68 responses from the participants. Unfortunately, four of the participants did not give their consent and submitted incomplete responses. Therefore, we analyzed 64 valid responses.
Table \ref{table:reasons-missing-code} shows our suggested reasons behind missing code and the participants' agreement levels with those reasons. According to our survey responses, most participants (i.e., 59.4\%) agree that users miss the required code snippets because they are unaware of whether their questions might need them (Table \ref{table:reasons-missing-code} (a)). Besides, more than half of the participants agree that users might struggle to decide which portion of code they should include in their questions (Table \ref{table:reasons-missing-code} (f)). However, they mostly disagree that users do not have example code ready when submitting their questions (Table \ref{table:reasons-missing-code} (b)). All these findings suggest that an automatic identification 
can support users to decide whether their questions require any code.

We also capture free-form responses from the participants regarding the reasons behind the missing code. 
Thirty participants submit one or multiple reasons. Table \ref{table:other-reasons-missing-code} summarizes the reasons behind missing code from the free-form responses. We found that laziness and the users' lack of knowledge about the implications of missing code are to blame.

\begin{tcolorbox}[colframe=black!50, colback=white,left=0pt,right=1pt,top=1pt,bottom=1pt, arc=0.5pt]

    \emph{\textbf{Summary of RQ2:}} According to the survey, organizations might restrict uploading code in a public Q\&A site, or there might be privacy/security concerns. However, most users are unaware of (a) whether their questions need code snippets to be answered appropriately and (b) the negative impact of missing code snippets.

\end{tcolorbox}
\vspace{-2mm}

\begin{table}[t]
\centering
\caption{Experimental Results}
\label{table:result-analysis}
\resizebox{3in}{!}{%
\begin{tabular}{l|l|l|c|c|c|c}
\hline
\multicolumn{1}{c|}{\textbf{Features}} & \multicolumn{1}{c|}{\textbf{Technique}} & \multicolumn{1}{c|}{\textbf{Category}} & \textbf{Precision} & \textbf{Recall} & \textbf{F1-Score} & \textbf{Accuracy} \\ \hline

\multirow{12}{*}{\RotText{\textbf{Keywords}}} & \multirow{2}{*}{{RF}} &  {Code}     & 81.3\%  & 77.9\%  & 79.6\%  & \multirow{2}{*}{80.0\%}  \\ \cline{3-6}
                                  & & {No Code}  & 78.9\%  & 82.0\%  & 80.4\%  &                         \\ \cline{2-7}
                             
& \multirow{2}{*}{{XGBoost}} & {Code}     & {83.8\%}  & 80.1\%  & {81.9\%}  & \multirow{2}{*}{{\textbf{82.3}\%}}     \\ \cline{3-6}
                                 & & {No Code}   & 81.0\%  & {84.5\%}  & {82.7\%}  &                         \\ \cline{2-7}
                                  
& \multirow{2}{*}{{ANN}} & {Code}         & 83.1\%  & 77.4\%  & 80.1\%  & \multirow{2}{*}{80.8\%}     \\ \cline{3-6}
                                 & & {No Code}   & 78.8\%  & 84.3\%  & 81.4\%  &                         \\ \cline{2-7}
                                  
& \multirow{2}{*}{{NB}}    & {Code}       & 74.1\%  & {87.9\%}  & 80.4\%  & \multirow{2}{*}{78.6\%}     \\ \cline{3-6}
                                 & & {No Code}   & {85.1\%}  & 69.3\%  & 76.4\%  &                         \\ \cline{2-7}
                                  
& \multirow{2}{*}{{KNN}}    & {Code}     & 76.8\%  & 64.1\%  & 69.9\%  & \multirow{2}{*}{72.4\%}     \\ \cline{3-6}
                                 & & {No Code}  & 69.2\%  & 80.7\%  & 74.5\%  &                         \\ \cline{2-7}
                                  
& \multirow{2}{*}{{SVM}}    & {Code}     & 83.0\%  & 78.5\%  & 80.7\%  & \multirow{2}{*}{81.2\%}     \\ \cline{3-6}
                                 & & {No Code}  & 79.6\%  & 84.0\%  & 81.7\%  &                         \\ \hline \hline

\multirow{12}{*}{\RotText{\textbf{POS Patterns}}} & \multirow{2}{*}{{RF}} &  {Code}     & 68.8\%  & 65.2\%  & 67.0\%  & \multirow{2}{*}{67.8\%}  \\ \cline{3-6}
                                  & & {No Code}  & 66.9\%  & 70.4\%  & 68.6\%  &                         \\ \cline{2-7}
                             
& \multirow{2}{*}{{XGBoost}} & {Code}     & 76.9\%  & 64.4\%  & 70.1\%  & \multirow{2}{*}{72.5\%}     \\ \cline{3-6}
                                 & & {No Code}   & 69.4\%  & 80.7\%  & 74.6\%  &                         \\ \cline{2-7}
                                  
& \multirow{2}{*}{{ANN}} & {Code}         & 73.8\%  & 65.5\%  & 69.4\%  & \multirow{2}{*}{71.1\%}     \\ \cline{3-6}
                                 & & {No Code}   & 69.0\%  & 76.8\%  & 72.7\%  &                         \\ \cline{2-7}
                                  
& \multirow{2}{*}{{NB}}    & {Code}       & 72.1\%  & {80.1\%} & {75.9\%}  & \multirow{2}{*}{{\textbf{74.6\%}}}     \\ \cline{3-6}
                                 & & {No Code}   & {77.6\%}  & 69.1\%  & 73.1\%  &                         \\ \cline{2-7}
                                  
& \multirow{2}{*}{{KNN}}    & {Code}     & 65.7\%  & 73.5\%  & 69.4\%  & \multirow{2}{*}{67.5\%}     \\ \cline{3-6}
                                 & & {No Code}  & 69.9\%  & 61.6\%  & 65.5\%  &                         \\ \cline{2-7}
                                  
& \multirow{2}{*}{{SVM}}    & {Code}     & {79.4\%}  & 60.5\%  & 68.7\%  & \multirow{2}{*}{72.4\%}     \\ \cline{3-6}
                                 & & {No Code}  & 68.1\%  &{ 84.3\%}  & {75.3\%}  &                         \\ \hline \hline

\multirow{12}{*}{\RotText{\textbf{\begin{tabular}{l}Code Element \& \\ Sentence Structure\end{tabular}}}} & \multirow{2}{*}{{RF}} &  {Code}     & 63.9\%  & 46.4\%  & 53.8\%  & \multirow{2}{*}{60.1\%}  \\ \cline{3-6}
                                  & & {No Code}  & 57.9\%  & 73.8\%  & 64.9\%  &                         \\ \cline{2-7}
                             
& \multirow{2}{*}{{XGBoost}} & {Code}     & 63.9\%  & 47.0\%  & 54.1\%  & \multirow{2}{*}{60.2\%}     \\ \cline{3-6}
                                 & & {No Code}   & 58.1\%  & 73.5\%  & 64.9\%  &                         \\ \cline{2-7}
                                  
& \multirow{2}{*}{{ANN}} & {Code}         & 64.0\%  & 51.1\%  & 56.8\%  & \multirow{2}{*}{{\textbf{61.2\%}}}     \\ \cline{3-6}
                                 & & {No Code}   & {59.3\%}  & 71.3\%  & 64.7\%  &                         \\ \cline{2-7}
                                  
& \multirow{2}{*}{{NB}}    & {Code}       & {80.0\%}  & 17.7\%  & 29.0\%  & \multirow{2}{*}{56.6\%}     \\ \cline{3-6}
                                 & & {No Code}   & 53.7\%  & {95.6\%}  & {68.8\%}  &                         \\ \cline{2-7}
                                  
& \multirow{2}{*}{{KNN}}    & {Code}     & 54.7\%  & {68.0\%}  & {60.6\%}  & \multirow{2}{*}{55.8\%}     \\ \cline{3-6}
                                 & & {No Code}  & 57.7\%  & 43.7\%  & 49.7\%  &                         \\ \cline{2-7}
                                  
& \multirow{2}{*}{{SVM}}    & {Code}     & 65.5\%  & 46.7\%  & 54.5\%  & \multirow{2}{*}{61.0\%}     \\ \cline{3-6}
                                 & & {No Code}  & 58.6\%  & 75.4\%  & 65.9\%  &                         \\ \hline \hline

\multirow{12}{*}{\RotText{\textbf{All}}} & \multirow{2}{*}{{RF}} &  {Code}     & 83.5\%  & 85.4\%  & 84.4\%  & \multirow{2}{*}{84.3\%}  \\ \cline{3-6}
                                  & & {No Code}  & 85.0\%  & 83.2\%  & 84.1\%  &                         \\ \cline{2-7}
                             
& \multirow{2}{*}{{XGBoost}} & {Code}     & 85.3\%  & 83.4\%  & 84.4\%  & \multirow{2}{*}{84.5\%}     \\ \cline{3-6}
                                 & & {No Code}  & 83.8\%  & 85.6\%  & 84.7\%  &                         \\ \cline{2-7}
                                  
& \multirow{2}{*}{{ANN}} & {Code}         & 83.3\%  & 81.5\%  & 82.4\%  & \multirow{2}{*}{82.6\%}     \\ \cline{3-6}
                                 & & {No Code}  & 81.9\%  & 83.7\%  & 82.8\%  &                         \\ \cline{2-7}
                                  
& \multirow{2}{*}{{NB}}    & {Code}       & 80.4\%  & {90.8\%}  & {85.3\%}  & \multirow{2}{*}{84.4\%}     \\ \cline{3-6}
                                 & & {No Code}  & {89.5\%}  & 77.9\%  & 83.3\%  &                         \\ \cline{2-7}
                                  
& \multirow{2}{*}{{KNN}}    & {Code}      & 78.2\%  & 63.3\%  & 69.9\%  & \multirow{2}{*}{72.8\%}     \\ \cline{3-6}
                                 & & {No Code}  & 69.1\%  & 82.3\%  & 75.2\%  &                         \\ \cline{2-7}
                                  
                                  
& \multirow{2}{*}{{SVM}}    & {Code}     & {86.5\%}  & 83.4\%  & 85.0\%  & \multirow{2}{*}{{\textbf{85.2}\%}}     \\ \cline{3-6}
                                 & & {No Code}  & 84.0\%  & {87.0\%} & {85.5\%}  &                         \\ \hline 
                               
\end{tabular}
}
\vspace{-3mm}
\end{table}

\subsection{Prediction Models to Identify the Questions Requiring Code Snippets (RQ3)}
\label{subsec-findings:predictionModels}

In this section, we evaluate the performances of our ML models and compare our models with baseline models.


\noindent \textbf{Evaluation of our Prediction Models.}
Table \ref{table:result-analysis} summarizes the evaluation details of our models. First, we develop simpler models to check the strength of each feature type. We find that keywords are the strongest features for discriminating between the questions that need and do not need code snippets. For example, keywords can identify the questions that need code snippets with the highest precision of 83.8\%, recall of 87.9\%, F1-score of 81.9\%, and overall accuracy of 82.3\%. Such statistics are 79.4\%, 80.1\%, 75.9\%, and 74.6\% when we train \& test our models with POS patterns only. We also see a performance drop in our models when we experiment with the remaining two features - code elements \& sentence structure.

We then combine all four types of features and analyze the performances of our models. As shown in Table \ref{table:result-analysis}, five models (out of six) can identify both question types (need code/do not need code) with more than 80\% precision. The highest precision (i.e., 86.5\%) was achieved by the SVM model when detecting questions that need code snippets. Such a finding suggests that our SVM model can succeed in 8 out of 10 cases, which is highly promising.
We also note similar results for the recall and F1-score. For example, the NB model achieves the highest recall (i.e., 90.8\%) and F1-score (i.e., 85.3\%) when detecting questions that need code snippets. However, SVM achieved the highest recall (i.e., 87\%) and F1-score (i.e., 85.5\%) for the questions that do not need code snippets. Such consistent results across multiple models with different learning strategies indicate the strength of our selected features. Overall accuracy is more than 82\% for all the models except KNN. However, SVM marginally outperforms the other models by achieving an overall accuracy of 85.2\%. The performance of KNN is somewhat lower than that of the others. 
Since KNN is a distance-based algorithm, its performance might be affected for high-dimensional features, which is the case for our dataset \cite{knnLimitations}. 

\begin{tcolorbox}[colframe=black!50, colback=white,left=0pt,right=1pt,top=1pt,bottom=1pt, arc=0.5pt]

    \emph{\textbf{Summary of RQ3:}} We develop six ML models using four text-based features. 
    Our models can identify the questions that need code snippets with promising precision (e.g., 86.5\%) and recall (e.g., 90.8\%).

\end{tcolorbox}
\vspace{-2mm}

\section{Threat to Validity} 
\label{sec:threatToValidity}

Threats to \emph{external validity} relate to the generalizability of a technique. We randomly selected 1,650 questions, analyzed them to extract features, and then developed our ML classifiers. However, we did not target any specific programming languages. We also extracted the dataset from the whole SO corpus of August 2023. Thus, our findings might be generalizable for SO. However, we cannot guarantee the same findings for other Q\&A sites.

Threats to \emph{internal validity} relate to experimental errors and biases \cite{tian2014automated}. We chose seventeen key phrases (e.g., add code) by manually analyzing the question comments. Then, we identify the target questions and construct our datasets based on those key phrases. However, those key phrases might be limited to extracting all the target questions and could lead to false positive samples. We thus further validate each of the selected questions involving two human annotators to avoid any false positives (Section \ref{subsec-mothodology:createDataset}). 
The agreement between the two annotators was almost perfect (i.e., {\large $\kappa$} $= 0.99$).
Another error might stem from the question editing time when the requested code is added. According to our investigation, a few of the edits went through reviews. We collected the edit approved time rather than the actual question editing time in those cases. However, in our dataset, less than 1\% edits undergo reviews. 
Thus, they might have a negligible impact on our overall findings.

Threats to \emph{construct validity} relate to the suitability of evaluation metrics and tests. We evaluate the performance of our models using four appropriate metrics (precision, recall, F1-score \& accuracy) \cite{beyer2018automatically,mondal2021early}. We use \emph{Mann-Whitney-Wilcoxon} test \cite{mcknight2010mann}, which is a widely used non-parametric test for evaluating the difference between two sample sets \cite{emse2021bmondal}. However, the significance level might suffer due to the limited size of the samples. We thus consider the effect size along with the \emph{p-value}. 
To see the correlation between two categorical variables, we also use the \emph{Chi-squared} test. This statistical test of independence works well when there is a  small number of categories ($\le 20$) \cite{chiSquareTest}. Thus, threats to construct validity might be mitigated.

Snowball sampling relies on referrals and may have a sampling bias. However, we also selected participants using an open circular approach and collected their responses anonymously. As shown in Table \ref{table:participants-summary}, our participants have diverse experiences and professions. Such diversity offers validity and applicability to our survey findings. However, any individual bias in the survey responses should be mitigated via a large sample of 64 participants.

\section{Related Work} 
\label{sec:relatedwork}

Several existing studies investigate whether including code snippets in the SO questions improves their quality or encourages faster answers \cite{asaduzzaman2013answering, squire2014bit, calefato2018ask, wang2018understanding, neshati2017early, baltadzhieva2015predicting, chua2015answers, yazdaninia2021characterization, qclassification, treude2011programmers}. Squire and Funkhouser~\cite{squire2014bit} analyze the role of source code in SO questions. According to them, a bit of code in the questions might encourage more positive votes on average. 
Wang et al.~\cite{wang2018understanding} and Calefato et al.~\cite{calefato2018ask} also suggest that code snippets could improve the quality of a question and encourage faster answers.
On the contrary, Baltadzhieva et al.~\cite{baltadzhieva2015predicting} investigate the impact of code snippets on questions and suggest that the presence of code snippets might hurt the chance of getting answers.
However, none of the above studies identify the questions requiring code snippets, which we focus on in our work.

Several studies utilize the presence or length of code snippets in their models to identify high-quality (or answered) questions \cite{neshati2017early, baltadzhieva2015predicting, chua2015answers, yazdaninia2021characterization}. In particular, they attempt to correlate high-quality or answerable questions and the presence of code snippets. According to their analysis, the presence of code snippets has both negative \cite{chua2015answers, baltadzhieva2015predicting} and positive \cite{calefato2018ask} effects on questions getting answers. However, missing the required could prevent the questions from getting their answers. On the other hand, including redundant code snippets could increase analysis overhead, which is not recommended. Our study analyzes the effect of missing but required code snippets in SO questions and demonstrates its negative implications. To support the SO users, we also design automated models to identify the questions that need code snippets.

Beyer et al.~\cite{beyer2018automatically} developed seven ML-based classifiers to identify seven categories (e.g., discrepancy) of SO questions. 
However, their study does not investigate or suggest which categories of questions need code snippets.
Ford et al.~\citep{ford2018we} deploy a month-long, just-in-time mentorship program to improve the quality of SO questions. Mentors guide novice users with informative feedback on their questions. Such mentorship reduces delays in getting answers. However, human mentorship is costly and, thus, challenging to sustain. Therefore, tool supports that can assist users to improve their questions by automatically identifying quality issues can be a sustainable solution. For example, a tool can interact with our models to examine whether a question misses required code snippets and guide users.
Chaparro et al.~\cite{chaparro2017detecting} attempt to detect missing information in bug reports utilizing discourse patterns. Our study partly overlapped with their methodology. However, our problem domain and context are different -- our study attempts to detect missing code snippets in SO questions.
To the best of our knowledge, the effect of missing but required code snippets in SO questions was not investigated before, and there exists no work to automatically detect such questions, which makes our work novel.

\section{Conclusion and Future Work}
\label{sec:conclusion}

Stack Overflow questions that discuss code-related issues require example code snippets to be answered appropriately. Unfortunately, users often miss required code snippets, which could prevent their questions from getting prompt and appropriate answers. This study analyzes the impact of missing required code in SO questions and designs a novel technique to detect them automatically. According to our investigation, only 23.8\% of questions get acceptable answers that miss code snippets as opposed to 61.4\% that include code during their submission. Interestingly, adding code snippets to questions upon request increases the chance of getting acceptable answers by 20\%. 28\% of questions remain unanswered when they lack the required code snippets. Confounding factors (e.g., user reputation) might affect questions receiving answers. Nevertheless, regardless of such factors, including necessary code snippets consistently encourages acceptable answers to questions with maximum delay. We surveyed 64 SO users, who mostly agreed that users need to be made aware of whether their questions need code snippets.
We thus develop six ML models to automatically identify the questions needing code snippets that can predict the target questions with the highest precision of 86.5\% 
We aim to -- (a) introduce tool support to the SO question submission system to detect questions that need code snippets during submission and (b) survey practitioners to evaluate its effectiveness in the future.

\smallskip
\noindent \textbf{Acknowledgment:}
This research is supported in part by the Natural Sciences and Engineering Research Council of Canada (NSERC) Discovery grants, and an NSERC Collaborative Research and Training Experience  (CREATE) grant, and two Canada First Research Excellence Fund (CFREF) grants coordinated by the Global Institute for Food Security (GIFS) and the Global Institute for Water Security (GIWS).

\balance


\bibliographystyle{unsrtnat}
\bibliography{reference}

\end{document}